\documentclass[letterpaper,twoside]{article}
\usepackage{calc,amsmath,amssymb,amsfonts}
\usepackage{comment}
\usepackage[LGR,T1]{fontenc}
\usepackage{natbib}
\usepackage[greek,english]{babel}
\usepackage{blindtext}
\usepackage{multicol}
\usepackage[font=small,labelfont=bf]{caption}
\newenvironment{Table}
  {\par\bigskip\noindent\minipage{\columnwidth}\centering}
  {\endminipage\par\bigskip}
\usepackage{xcolor,longfbox,multicol,fancyhdr}
\usepackage[top=0.3in,bottom=0.3in,inner=1in,outer=1in,includehead,head=0.2in,headsep=0.15in]{geometry}
\usepackage{array,supertabular,hhline,enumitem,hyperref}
\usepackage{multirow}
\usepackage{wrapfig}
\usepackage{caption}
\hypersetup{colorlinks=true,allcolors=black}
\usepackage[pdftex]{graphicx}
\usepackage{pdfpages} 
\usepackage{afterpage} 
\usepackage{tikz}
\usepackage{booktabs} 
\usepackage{titlesec} 
\usepackage{adjustbox}
\titleformat{\section}
  {\normalfont\Large\bfseries}{\thesection}{0.5em}{}

\titleformat{\subsection}
  {\normalfont\large\bfseries}{\thesubsection}{0.5em}{}

\fancypagestyle{Standard}{\fancyhf{}
  \fancyhead[LO]{\color{lightgray}\rmfamily\bfseries\itshape GENEVIC}
  \fancyhead[LE]{\color{lightgray}\rmfamily\bfseries\itshape A.Nath et al.
  \begin{tikzpicture}
\path[line width=0.0071in,draw=black] (0,0) -- (17.780,0);
\end{tikzpicture}
}

  \fancyfoot[L]{}

}
\fancypagestyle{FirstPage}{\fancyhf{}
  \fancyhead[L]{}
  \fancyfoot[L]{}

}
\title{GENEVIC}
\author{anindita.nath}
\date{2024-01-15}
\begin{document}
\clearpage
\pagestyle{Standard}
\thispagestyle{FirstPage}
{\raggedleft\selectlanguage{english}
\textit{Bioinformatics}, 2024
\par}

{\raggedleft\selectlanguage{english}
doi: 10.1093/bioinformatics/xxxxx
\par}

{\raggedleft\selectlanguage{english}
Submitted: 29 March 2024
\par}

{\raggedleft\selectlanguage{english}
Applications Note
\par}

\begin{flushleft}
\tablefirsthead{}
\tablehead{}
\tabletail{}
\tablelasttail{}
\begin{supertabular}{m{6.5in}}
\noalign{\hrule height 1pt} 
\noalign{\vskip 2pt} 
{\selectlanguage{english}\Large \itshape Subject Section}\\
\noalign{\vskip 2pt} 
{\selectlanguage{english}\Large\bfseries GENEVIC: GENetic data Exploration and Visualization via Intelligent interactive Console}\\

{\selectlanguage{english} \Large Anindita Nath\textsuperscript{1}, Savannah Mwesigwa\textsuperscript{1}, Yulin Dai\textsuperscript{1}, Xiaoqian Jiang\textsuperscript{2} and Zhongming Zhao\textsuperscript{1,3,*}  }

{\selectlanguage{english} \large\textsuperscript{1}Center for Precision Health, McWilliams School of Biomedical Informatics, The University of Texas Health Science Center at Houston, Houston, TX 77030, USA}

{\selectlanguage{english} \large\textsuperscript{2} Department of Health Data Science and Artificial Intelligence, McWilliams School of Biomedical Informatics, The University of Texas Health Science Center at Houston, Houston, TX 77030, USA}

{\selectlanguage{english} \large\textsuperscript{3} MD Anderson Cancer Center, UTHealth Graduate School of Biomedical Sciences, Houston, TX 77030, USA}

{\selectlanguage{english} *To whom correspondence should be addressed.}


{\selectlanguage{english} Received on 03-29-2024}\\
\noalign{\vskip 5pt} 

{\selectlanguage{english}\bfseries\upshape Abstract}

{\selectlanguage{english} \textbf{Summary:} The vast generation of genetic data poses a significant challenge in efficiently uncovering valuable knowledge. Introducing GENEVIC, an AI-driven chat framework that tackles this challenge by bridging the gap between genetic data generation and biomedical knowledge discovery. Leveraging generative AI, notably ChatGPT, it serves as a biologist's 'copilot'. It automates the analysis, retrieval, and visualization of customized domain-specific genetic information, and integrates functionalities to generate protein interaction networks, enrich gene sets, and search scientific literature from PubMed, Google Scholar, and arXiv, making it a comprehensive tool for biomedical research. In its pilot phase, GENEVIC is assessed using a curated database that ranks genetic variants associated with Alzheimer's disease, schizophrenia, and cognition, based on their effect weights from the Polygenic Score (PGS) Catalog, thus enabling researchers to prioritize genetic variants in complex diseases. GENEVIC's operation is user-friendly, accessible without any specialized training, secured by Azure OpenAI's HIPAA-compliant infrastructure, and evaluated for its efficacy through real-time query testing. As a prototype, GENEVIC is set to advance genetic research, enabling informed biomedical decisions.}
\\
{\selectlanguage{english} \foreignlanguage{english}{\textbf{Availability and implementation:}}\foreignlanguage{english}{ GENEVIC is publicly accessible at \href{https://genevic-anath2024.streamlit.app/}{https://genevic-anath2024.streamlit.app}. The underlying code is open-source and available via GitHub at \href{https://github.com/anath2110/GENEVIC.git}{https://github.com/anath2110/GENEVIC.git}
}}\\
{\selectlanguage{english} \foreignlanguage{english}{\textbf{Contact:}}\foreignlanguage{english}{ Zhongming.Zhao@uth.tmc.edu}}\\
{\selectlanguage{english} \foreignlanguage{english}{\textbf{Supplementary
information:}}\foreignlanguage{english}{\textcolor{blue}{ }}\foreignlanguage{english}{ Supplementary data
}\foreignlanguage{english}{\textcolor{black}{are available at
}}\foreignlanguage{english}{\textit{\textcolor{black}{Bioinformatics }}}\foreignlanguage{english}{\textcolor{black}{ online and also at \href{https://github.com/anath2110/GENEVIC\_Supplementary.git}{https://github.com/anath2110/GENEVIC\_Supplementary.git}}}}\\
\noalign{\vskip 1.5pt} 
\noalign{\hrule height 1pt} 
\end{supertabular}
\end{flushleft}

\begin{multicols}{2}
\section{Introduction}
\begin{figure*}[!ht]
\centering
\includegraphics[width=1.0\linewidth]{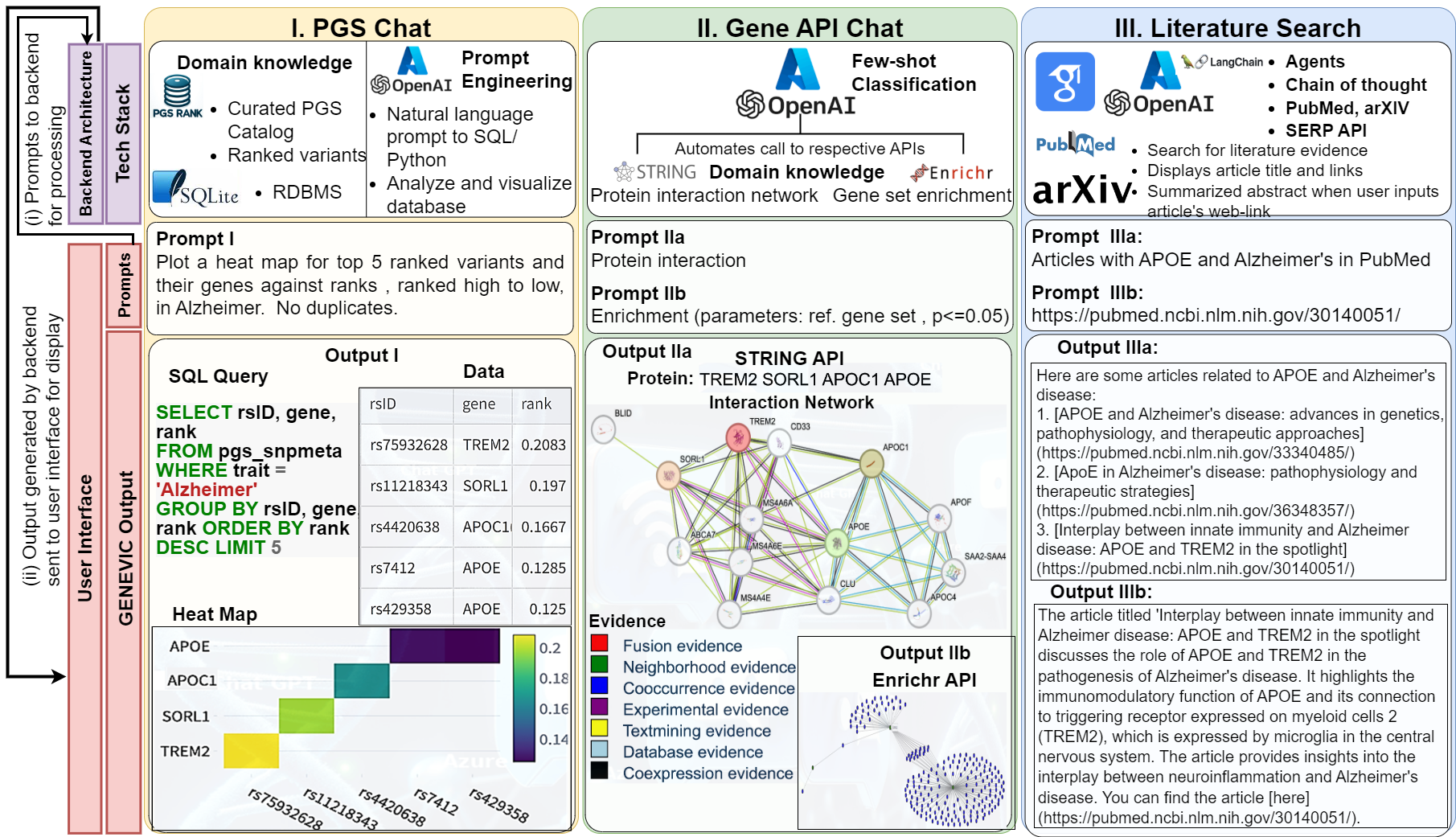}

\caption{Workflow overview between the front-end user interface and the back-end architecture of GENEVIC: (i) User prompts from the interface are sent to the back-end for processing and then, (ii) Generated output is sent back to the user. The backend comprises common AI services, supported by data tools exclusive to each of the three functionalities: PGS Chat (I), GENEAPI Chat (II), and Literature Search (III). The user prompts: Prompt Ia, Prompt IIa, Prompt IIb, Prompt IIIa and Prompt IIIb generate GENEVIC outputs: Output Ia, Output IIa, Output IIb, Output IIIa and output IIIb, respectively.}
\label{fig:combofig}
\end{figure*} 
Generative AI, notably Chat GPT and GPT-4 \citep{OpenAI2023}, excels in a wide array of natural language processing tasks, radically altering knowledge access and processing. Despite the surge in biomedical knowledge and tools, leveraging these advancements fully often requires deep domain and data science expertise, a challenge for many researchers. Additionally, these AI models, while capable of generating seemingly accurate outputs, can sometimes produce fictitious information, known as the 'hallucination effect' \citep{Rohrbach2018, Xiao2021}, complicating their reliability in specialized fields. To address this, integrating large language models with domain-specific databases or websites can enhance response accuracy in areas such as genomics \citep{Jin2023}, health management \citep{Issom2021}, and literacy \citep{Mokmin2021}. \\ Here, we introduce GENEVIC, an intelligent interactive console powered by Azure Open AI’s generative AI for genetic data exploration and visualization. To demonstrate its utility, we focus on user engagement with the information from Polygenic Score (PGS) Catalog \citep{Lambert2021}, a web-based database of polygenic risk scores (\href{https://www.pgscatalog.org}{https://www.pgscatalog.org}). This database is mainly driven by users’ deposition of published PGS values that include the variants, alleles, and weights in various phenotypes (e.g. complex diseases), providing a natural resource for GENEVIC to extract prioritized disease-relevant variants. Accordingly, GENEVIC aids in extracting pivotal genetic variants related to diseases, paving the way for building an extensive map of variant-gene-trait associations.\\
PGS quantifies an individual’s genetic predisposition to a specific disease or trait by aggregating genome-wide genotype effects, derived from genome-wide association studies (GWAS) data \citep{Choi2020}. While PGS has been shown promising to evaluate disease risk recently, \citet{Lambert2021} highlighted the lack of standardized reporting practices, which hampers PGS research progress. The PGS Catalog addresses this by providing a comprehensive repository of published PGS values, complete with essential metadata for accurate application and assessment, fostering reproducibility. Yet, the PGS Catalog’s static format limits interactive exploration, impeding the in-depth analysis of its rich dataset. Importantly, it lacks the capability for data consolidation, preventing the integration of various analyses such as network analyses, literature searches, and visualization for the related traits. This gap underscores the need for interactive tools like GENEVIC that enable dynamic engagement with the PGS Catalog, thereby enhancing data mining and the discernment of intricate patterns. GENEVIC not only facilitates detailed analysis and informed decision-making for researchers but also democratizes access to genetic research insights, allowing educators and laypersons to delve into variant functions and hypothesize to advancing the exploration of genetic knowledge.\\
GENEVIC enhances AI by integrating bioinformatics APIs like STRING and ENRICHR, and supports literature searches from major sites, enriching prioritized variants with specialized knowledge. This integration streamlines research, aiding in the identification of genetic markers and pathways. This a novel prototype of an end-to-end solution that incorporates knowledge, data, and agents into one single portal for a specific domain and thus, broadening access to domain expertise, simplifying data for diverse research backgrounds and promoting multidisciplinary collaboration.
\section{Overview of GENEVIC}
\subsection{Functionality}
GENEVIC is a user-friendly intelligent interactive console (or interface) implemented with computational methods to facilitate genetic research.\\
At its core, the PGSChat component allows researchers to input specific genetic data points, such as single nucleotide polymorphisms (SNPs, including rsIDs and genomic coordinates), gene symbols, and disease or phenotype names as part of prompts, to retrieve relevant information from the PGS rank database, which is a SQLite database that houses variant rankings derived from the PGS Catalog. This interface connects to a GeneAPI Chat interface, utilizing Enrichr and STRING web APIs for enrichment analysis using popular gene set libraries, and generating gene-gene interaction networks as well as visualizing the corresponding network graphs, respectively, for a set of genes provided as input. Concurrently, the Literature Search component streamlines the search and retrieval of scientific literature from PubMed, Google Scholar, and arXiv, for a given search query/prompt. Additionally, this particular component can retrieve the abstracts of an article for a given link to that article.

\subsection{Implementation} 
GENEVIC's architecture (Figure \ref{fig:combofig}) is split between a responsive front end, facilitating user interaction and visualization, and a robust back end, where data processing and API integration occur by harnessing and integrating the power of generative AI. Test cases showcasing sample prompts and corresponding GENEVIC's outputs for each of the three task components are shown in Figure \ref{fig:combofig}I, Figure \ref{fig:combofig}II and Figure \ref{fig:combofig}III, respectively.\\
The frontend/user interface was developed using Streamlit. The backend technology of our application is anchored by Azure OpenAI's ChatGPT 3.5 (or GPT-4, if available), a state-of-the-art generative AI model. 
\begin{Table}
  \begin{tabular}{p{2.4cm}p{1.2cm}p{3.2cm}}
    Phenotype (mapped trait)& \#PGS Files & SNP info \\
    \hline
    Alzheimer &23 & w:[(-0.95)-1.64]; m:0.0013; md: 5.82e-07; sd: 0.0390\\
    schizophrenia & 5 & w:[(-0.04)-0.05]; m:5.61e-05; md:8.36e-06; sd:0.0018 \\
cognition & 5 & w:[(-3.60)-1.68]; m:(-0.0005); md:-7e-08; sd:0.0321 \\
\hline
  \end{tabular}
 \captionof{table}{Summary of statistics of the phenotypes used to develop the PGS rank database. Phenotype denotes disease or trait name, \#PGS Files denote the count of PGS files for each PGS ID corresponding to the phenotype, SNP info refers to the aggregated information regarding the SNPs or variants in each PGS file. w is the weight range, m, md, sd denote the mean, median, and standard deviation of weights, respectively.}
 \label{tab:pgsdb}
\end{Table}
Generative AI is pivotal in this application, utilizing prompt-based few-shot learning to classify user prompts for smooth navigation and function execution. It efficiently translates English prompts into SQL and Python codes for querying the PGS rank database and creating intuitive data visualizations. This technology effectively connects to web APIs and enrichment tools with minimal prompts and simplifies literature searches across websites, demonstrating ChatGPT's versatility in managing various tasks and queries.
\subsection{ PGS rank database}
Our study harnessed PGS files from PGS Catalog (Release: August 4\textsuperscript{th}, 2023)  harmonized to human reference genome (GRCh38 build), creating a local database for easier access. We extracted trait-specific data using R's "Quincunx" package \citep{Magno2021}, querying the catalog's REST API with the mapped traits (ontology) as listed in Table \ref{tab:pgsdb}. To ensure consistency amid the dataset's diversity, we focused on essential columns: effect allele, effect weight, and SNP rsID/SNP coordinates, and harmonized variant labeling discrepancies. We merged phenotype-specific PGS files into a single dataset using R's "dplyr"\citep{dplyr23} and "bind\_rows" methods and treated missing values as `NA'.\\
We used the Dowdall method, \citep{Fraenkel2014}, an alternative Borda method,  for rank aggregation of the variants across multiple PGSs based on the absolute values of their effect weights, and annotated the variants using ANNOVAR\citep{ANNOVAR2010}. The rank aggregation step involved assigning the reciprocal of ranks (RR) to each variant. This means that the top rank receives a RR of 1, the second rank receives a RR of 1/2, the third rank receives a RR of 1/3, and so forth. For each variant query, we calculate the mean of the reciprocals of its ranks (MRR) across multiple PGSs. In cases where weights are unavailable, we assign an MRR of 0. The aggregated ranking of variants by effect weight thus assigns a higher score to variants that have a higher effect weight consistently across multiple PGSs and streamlines querying of trait-relevant top-ranked variants, offering a means to assess the functionality of the PGS Chat feature.\\
This database was developed using SQLite for its lightweight efficiency. Users can optionally switch to any custom SQL server-hosted database.
\subsection{Installation}
GENEVIC is accessible freely on Streamlit community cloud at \href{https://genevic-anath2024.streamlit.app}{https://genevic-anath2024.streamlit.app}, ready to use without installation. Due to Streamlit's 1 GB data limit, the PGS rank database is pared down to the top 100 genes each for Alzheimer's, schizophrenia, and cognition, adequate for this pilot project but is set for expansion in future cloud deployments with higher capacity. It's also runnable locally, as detailed in the installation guide. \href{http://tinyurl.com/InstalaltionGuide}{http://tinyurl.com/InstalaltionGuide}, with Python 3.10 or higher version as the only pre-requisite.\\
Users are required to have an active subscription with Azure OpenAI (refer \href{http://tinyurl.com/AzureOpenAIInstructions}{http://tinyurl.com/AzureOpenAIInstructions}), deploy ChatGPT or GPT-4 model, and then enter their account details in the `Settings' section of  GENEVIC's interface.\\

\subsection{Test cases}
GENEVIC's efficiency was evaluated through simulated real-world research tasks. We used ChatGPT 3.5-16k model for our test purposes. Its results for these simulations were compared with outcomes from direct website queries or traditional analysis methods. Notably, GENEVIC translates natural language prompts into SQL or Python code, allowing users to verify result accuracy by running this code in respective IDEs. Documented at \href{http://tinyurl.com/Test-Runs}{http://tinyurl.com/Test-Runs}, each test confirmed GENEVIC's capability to enhance research efficiency, offering rapid data access and effective visualization, vital in fast-paced research environments.\\While GENEVIC may occasionally struggle with precise result retrieval from the PGS rank database due to vague prompts, such challenges can be overcome by refining the queries. For example, enhancing the specificity of the prompt from "Show me the top 10 ranked genes in Alzheimer, top to bottom," to "If duplicate, show once," not only refines the search but also significantly improves the accuracy and relevance of the results, demonstrating GENEVIC's adaptability in navigating complex datasets.

\section{Discussion}
\subsection{Principle contributions}
GENEVIC showcases the capability of cutting-edge generative AI to unify and streamline access to, navigation of, and automate analysis of biomedical databases and external web APIs, marking a significant advancement. This platform with an intuitive and user-friendly interface, operates seamlessly, without requiring users to have specific technical or biomedical knowledge or training. It leverages standardized databases to ensure accuracy and minimize the risk of AI-generated misinformation, or "hallucination." Additionally, GENEVIC allows users to customize data sources and ensures data security through Azure OpenAI's HIPAA-compliant infrastructure, protecting sensitive clinical data.
\subsection{Limitations and future directions}
This pilot framework, GENEVIC, is only in its nascent stage, designed to evolve with advancements in ChatGPT and related technologies. Currently, it demonstrates capabilities using a limited PGS rank database with data for only three phenotypes and a basic approach to ranking variants via PGS effect weights, highlighting the need for comprehensive data, robust weighing scheme and consideration of various factors such as ethnic background, genotype data, specific PGS scoring techniques, and the degree of sample overlap across PGS datasets for each phenotype, among others.\\
Future enhancements will broaden the database scope, integrate additional biomedical web APIs into GeneAPI Chat, and enhance Literature Search functionalities by enabling auto-extraction of deeper insights. Plans to introduce automated predictive modeling using generative AI promise to significantly boost GENEVIC's functionality.\\ Thus, this innovative tool not only streamlines research workflows but also sets the stage for equipping future researchers with sophisticated, data-driven tools in genomics and biomedical research.

\section*{Acknowledgement}
We thank UT Health Science Center at Houston's technical team for their Azure Open AI setup support and colleagues from the Bioinformatics and Systems Medicine Laboratory and the Department of Health Data Science and Artificial Intelligence for their insightful feedback on this project.
\section*{Funding}
ZZ and XJ were partially supported by the National Institutes of Health (NIH) grant [U01AG079847]. ZZ was also partially supported by NIH grants [R01LM012806, R01LM012806-07S1, and R01DE030122]. We thank the technical support from the Cancer Genomics Core funded by the Cancer Prevention and Research Institute of Texas (CPRIT) [RP180734].  XJ is a CPRIT Scholar in Cancer Research [RR180012], and was also partially supported by Christopher Sarofim Family Professorship, UT Stars award, UTHealth startup, and the NIH grants [R01AG066749, R01AG082721, R01AG066749, U24LM013755, U01TR002062, U01CA274576, and U54HG012510]. The sponsors had no role in study design, data collection and analysis, decision to publish, or manuscript preparation.
\section*{Supplementary data}\label{sec:supmat}
 Supplementary data are available at Bioinformatics online and also at
{\scriptsize \href{https://github.com/anath2110/GENEVIC\_Supplementary.git}{https://github.com/anath2110/GENEVIC\_Supplementary.git}} 
\section*{Data availability}
Data supporting the findings of this study are available in the supplementary materials online and at \href{https://github.com/anath2110/GENEVIC.git}{https://github.com/anath2110/GENEVIC.git}
\section*{Conflict of Interest}
None declared.
\bibliographystyle{plainnat} 
\bibliography{references} 

\end{multicols}

\end{document}